\documentclass{article}

\usepackage{multicol}
\usepackage{fontspec}

\usepackage{microtype}

\usepackage[margin=1in]{geometry} 
\usepackage{graphicx}
\usepackage{floatrow}
\usepackage{adjustbox}

\usepackage[labelformat=simple]{caption}
\usepackage[below]{placeins}

\usepackage{subfig}
\usepackage{pdflscape}
\usepackage{multicol}
\usepackage{booktabs}
\usepackage{supertabular}
\usepackage{array}
\usepackage{enumitem}
\usepackage{appendix}
\usepackage{titling}
\usepackage[noblocks]{authblk} 
\usepackage[symbol*,multiple]{footmisc} 
\setfnsymbol{chicago} 

\usepackage{mathtools}
\usepackage{amssymb}
\usepackage{mathabx}
\usepackage[amsmath, amsthm, thmmarks, hyperref]{ntheorem}
\usepackage{stmaryrd}
\SetSymbolFont{stmry}{bold}{U}{stmry}{m}{n}

\usepackage[hidelinks]{hyperref}
\usepackage{xcolor}
\hypersetup{
    colorlinks,
    linkcolor={red!50!black},
    citecolor={blue!50!black},
    urlcolor={blue!80!black}
}

\usepackage[capitalize, noabbrev]{cleveref}
\crefname{equation}{eq.}{eqs.}
\crefname{appendix}{Supplementary Information}{Supplementary Information}

\usepackage[bibstyle=numeric-comp,citestyle=numeric-comp,uniquename=false,url=false,isbn=false,eprint=false,date=year,backend=biber,natbib,sorting=none,giveninits=true]{biblatex}

\renewcommand{\cite}{\autocite}
\renewcommand{\citep}{\autocite}

\renewbibmacro{in:}{\ifentrytype{article}
    {}
    {\bibstring{in}\printunit{\intitlepunct}}
}

\DeclareFieldFormat{doi}{%
  \textsc{doi}\addcolon\space
  \ifhyperref
    {\href{https://doi.org/#1}{\nolinkurl{#1}}}
    {\nolinkurl{#1}}%
}

\newcommand{\prob}[1]{\mathbb{P}\left[#1\right]}
\newcommand{\cc}{\boldsymbol{c}}
\newcommand{\mm}{\boldsymbol{m}}
\newcommand{\stirling}[2]{\genfrac{[}{]}{0pt}{}{#1}{#2}}

\newcommand{\probConfig}[1]{\mathbb{P}_{\textcolor{black}{\text{config}}}{\left[#1\right]}}
\newcommand{\probRealiz}[1]{\mathbb{P}_{\textcolor{black}{\text{realiz}}}{\left[#1\right]}}

\newcount\n
\def\tablebody{}
\makeatletter
\loop\ifnum\n<100
        \advance\n by1
        \protected@edef\tablebody{\tablebody
                \textbf{\number\n.}& shortText
                \tabularnewline
        }
\repeat

\makeatletter
\let\mcnewpage=\newpage
\newcommand{\TrickSupertabularIntoMulticols}{%
  \renewcommand\newpage{%
    \if@firstcolumn
      \hrule width\linewidth height0pt
      \columnbreak
    \else
      \mcnewpage
    \fi
  }%
}
\makeatother

\title{Quo nomine vis vocari? Random-copying, conformity, \\ and anti-conformity in the temporal sequence of papal names}

\author[1,$*$]{Egor Lappo}
\author[1]{Noah A.\ Rosenberg}

\affil[1]{Department of Biology, Stanford University}

\affil[$*$]{Email: \href{mailto:elappo@stanford.edu}{\texttt{elappo@stanford.edu}}}

\date{\today}
\begin{document}

\begin{titlingpage}

\maketitle


\noindent \textbf{Abstract.} The study of cultural evolution seeks to understand the processes by which behavioral variants are chosen in human cultures over time. The selection of new popes, each of whom chooses a papal name---typically reusing previous names in reference to previous popes---is among the longest ongoing cultural processes taking place in a single human institution. Here, we test the papal naming sequence against predictions of models developed in population genetics and stochastic processes---Ewens sampling theory and the Chinese restaurant process---which in the case of papal names amount to randomly copying an existing name in proportion to its frequency, with the possibility of innovation. Although papal name choices are careful individual decisions, we find that the sequence of papal names often accords with a random-copying model, with long periods in which a model of conformity in name choices more closely fits the name sequence and other periods in which anti-conformity provides a closer fit. Hence, despite the consideration that enters into individual choices, aggregate cultural behavior in a 2000-year old human process can potentially be described with simple laws. Our methods can potentially be used to deepen quantitative analyses of conformity and anti-conformity in other cultural processes that involve discrete cultural variants.

\medskip


\end{titlingpage}

\section{Introduction}\label{sec:intro}

Cultural evolutionary studies often investigate human choices of behavioral variants over time~\cite{henrich_evolution_2003, laland_social_2004, steele_ceramic_2010, premo_cultural_2014, kandler_analysing_2019, newberry_measuring_2022, lappo_cultural_2023, beheim_strategic_2014, beheim_opening_2025}. For example, Steele et al.~\cite{steele_ceramic_2010} examined Bronze Age Hittite ceramic bowls, studying changes in bowl features over hundreds of years. Newberry \& Plotkin~\cite{newberry_measuring_2022} investigated changes in the frequencies of baby names in the United States in the 20th century. Lappo et al.~\cite{lappo_cultural_2023} evaluated the move choices that professional chess players have made over the last several decades in specific board positions. In each of these cases, statistical tests were used to assess the fit of a cultural process to a model in which cultural variants are selected at random.

In the Catholic church, each newly elected pope chooses a name that is associated with his pontificate. \emph{Quo nomine vis vocari}, he is asked~\cite{conclave_explainer}: by what name do you wish to be called? The choice of name generally carries symbolic meaning and is carefully considered, as is illustrated in a quotation from a character in the 2025 film \emph{Conclave}: ``Every cardinal, deep down, has already chosen the name by which he would like his papacy to be known.'' The name usually references previous popes; for example, the current Pope Leo XIV cited his desire to continue in the path of Leo XIII (pontificate 1878-1903) \cite{vatican_leo}. 

Like the examples of ceramic bowls~\cite{steele_ceramic_2010}, baby names~\cite{newberry_measuring_2022}, and chess moves~\cite{lappo_cultural_2023}, sequential choices of pontifical names represent a human cultural process unfolding over time. Indeed, due to the $\sim$2000-year length of papal history in the Catholic church, this continuing process is perhaps one of the human cultural processes of longest duration involving specific identified individuals and a single institution. 

Examining the names of the 265 popes (Table \ref{table:popes}), we can observe a number of phenomena reminiscent of other cultural evolutionary settings. First, as in the bowl example~\cite{steele_ceramic_2010}, the popes possess a high diversity of names, with 81 distinct names used over the $\sim$2000 years of the papacy. Second, as is often the case with chess moves~\cite{lappo_cultural_2023}, no single name predominates, with the highest-frequency name having only 21 instances. Third, like baby names~\cite{newberry_measuring_2022}, although most new variants copy a previously used variant, novelty continues to be possible. The recent Pope Francis, though unusually for modern times, chose a novel papal name, not referencing a previous pope, instead referencing Saint Francis of Assisi \cite{francis}. 

Do the sequence and distribution of papal names follow systematic rules? We use the record of papal names as a setting for long-term analysis of human cultural behavior. Each papal name choice can be viewed as the result of a  careful individual decision. Nevertheless, we find that the name sequence over the \textasciitilde2000 years of the papacy accords with models based on the Ewens sampling theory in population genetics~\cite{ewens_sampling_1972, crane_ubiquitous_2016, tavare_lectures_2004} and the Chinese restaurant process in stochastic process theory~\cite{aldous_exchangeability_1985, pitman_exchangeable_1995, arratia_logarithmic_2003, tavare_magical_2021}---in which each pope randomly copies an existing name in proportion to a simple function of its frequency, allowing for new names: new ``mutations.''  

\section{Results}\label{sec:results}

We assembled a data set consisting for each of 265 popes of the papal name and the start date for that pope's term (Table \ref{table:popes}) \citep{annuario}. We also tabulated the multiplicity and most recent date of occurrence for each of the 81 distinct names (\cref{fig:data}). Although each pope technically has a unique binomial signified by the verbal name and the ordinal number counting popes with that name up to and including that pope (e.g.~Leo I, \ldots, Leo XIII, Leo XIV), in practice, the active pope is distinguished from other popes simply by being the current pope, so that the ordinal suffix is not generally needed in his identifier. Therefore, in counting appearances of a ``name,'' we tabulate counts of the verbal name rather than the (unique) binomials.

Using Ewens sampling theory, we examined the fit to the frequency distribution of names of a model in which each pope copies the name of a previous pope, chosen at random, with allowance for new name choices.  Informally, the selection of a papal name by a new pope often refers to previous popes with the same name, so that if all previous popes are treated as equiprobable as the basis for name-copying, among reused names, the probability that a specific name is chosen is proportional to the number of its previous uses. 

The Ewens distribution and Chinese restaurant process provide a mechanistic model for describing data unfolding over time according to a specified rule in which, conditional on a name choice not being a ``mutation,'' each choice randomly copies an existing name in linear proportion to its frequency. That is, under the random-copying model that results in the Ewens distribution, each draw has a novel type with probability proportional to ``mutation'' rate $\theta$, or an existing type with probability proportional to $i$ if the type has occurred $i$ times. In the setting of papal names, a mutation represents the selection of a name that has never been used, such as the selection of the name ``Francis'' in 2013. The probability that an existing name is chosen is proportional to its number of previous uses, e.g.~13 upon the ascent of Leo XIV in 2025. 

\subsection{Fit of the random-copying model}

To assess the fit of the observed counts of papal names to the Ewens distribution, we performed Slatkin's exact test \citep{slatkin_exact_1994, slatkin_correction_1996}. This test checks if the frequency distribution of names is unusual at significance level $\alpha$ by evaluating if the sum of the probability of the observed name frequency vector $\cc_0$ and the probabilities of all possible name frequency vectors with probability less or equal to than that of $\cc_0$, conditional on the number of distinct names, lies in one of the tails $[0,\alpha/2)$ or $(1-\alpha/2,1]$ (Methods). High values of the tail probability $P_E$ of Slatkin's exact test reflect conformity, and low values represent anti-conformity \citep{steele_ceramic_2010,premo_cultural_2014}.

For each pope, we conducted the test on the frequency vector for which that pope was the last one tabulated. For the first 43 popes, only a single frequency vector is possible conditional on the number of distinct names; as this vector has probability 1, the exact test is trivial ($i$ singleton names if the last pope tabulated is the $i$th pope, $1 \leq i \leq 23$; $i-2$ singleton names and one doubleton name if the last pope is the $i$th pope, $24 \leq i \leq 43$). The first nontrivial configuration is $i=44$, considering popes from Peter to Sixtus III, as two frequency vectors are possible for 44 popes and 42 distinct names: 41 singletons and 1 tripleton, and 42 singletons and 2 doubletons. 

\cref{fig:exact test}A shows the tail probability $P_E$ of Slatkin's exact test. With significance level $\alpha=0.05$, considering popes $i=44$ to 265, the tail probability lies below the $0.975$ cutoff for 83 popes, so that the random-copying model is not rejected. It lies above the $0.975$ cutoff for 139 popes, rejecting random-copying in the upper tail that reflects conformity in name choice. Throughout the entire duration of the history of papal name selection, the full sequence of papal name choices to date fits random-copying or rejects random copying in the upper tail corresponding to conformity.

In addition to computing the tail probability $P_E$ for the exact test, we rely on the fact that the number of distinct names in the frequency vector is a sufficient statistic for the ``mutation'' parameter $\theta$ to obtain a maximum likelihood estimate $\hat \theta$ (\cref{fig:exact test}B). $\hat \theta$ is high for the earliest configurations, when the probability that a new pope used a papal name that had never previously been used was high. This parameter then decreases, settling at $\hat \theta \approx 40$ for the last 500 years. Because the probability under the random-copying model is $\theta/(\theta+i-1)$ that the $i$th pope chooses a novel name, $\theta \approx 40$ means that prior to Leo XIV, the model would have predicted with probability $40/(40+265-1) \approx 13\%$ that the new pope would adopt a novel rather than an existing name. Under the model, a similar probability applies to the pope who will follow Leo XIV.

\subsection{Model fit in temporal windows}

We next conducted a moving-window analysis to assess name choices in particular periods. For this analysis, we used a modification of the Slatkin exact test that we call the \emph{conditional} Slatkin exact test (Methods), in which a subset of name choices is tested for agreement with random-copying, conditional on all prior names being held fixed. In particular, we tested sequences of name choices for popes selected within 200-year moving windows (e.g.~$[1425,1624]$) conditional on all previous names (e.g.~popes selected in 1424 or earlier). In addition, we performed an accompanying conditional maximum-likelihood estimation of $\theta$ to measure the mutation rate within temporal windows (Methods).

The conditional test finds that in most temporal windows, papal name selection accords with the random-copying model or rejects random copying with high values of $P_E$ corresponding to conformity. However, it identifies one period in which the tail probability is particularly low (\cref{fig:window}A). For the 34 windows of length 200 years ending 1223 to 1256, the test rejects random copying at the $\alpha=0.05$ significance level in 27 of the 34 years, so that for example, in $[1057,1256]$, random copying is rejected with a probability that corresponds to anti-conformity ($P_E < 0.025$). The tail probability also falls as low as 0.232 for the window ending in 1589, but it does not approach 0.025.

In \cref{fig:window}B, the conditional estimate of $\theta$ in temporal windows decreases over time in a similar manner to the unconditional estimate in \cref{fig:exact test}B. $\hat \theta$ reaches zero for windows $[913,1112]$ to $[1778,1977]$, as no novel names were introduced in these windows. $\hat \theta$ has increased in the most recent windows due to the novel names John Paul (1978) and Francis (2013).

One way to view a deviation from the random-copying model is in terms of ``evenness'': the frequency vector could be more ``even'' than is predicted by the model, with low-frequency names chosen more often than expected from the Ewens distribution, or more ``uneven,'' with high-frequency names instead chosen more often than expected. To examine these possibilities, \cref{fig:window}C plots the diversity of the papal names within temporal windows, as measured by the Shannon entropy of the name set. For the period in the 11th-13th centuries during which the conditional test rejects random copying, the entropy displays a peak, suggesting that the frequency vector became more even than expected. Although random-copying holds during the most recent windows (\cref{fig:window}A), \cref{fig:window}C also shows the lowest entropy during the 1900s, likely attributable in part to the fact that for the period 1740-1939, 7 of 14 popes shared a single name, ``Pius.'' 

\subsection{Fit of conformity and anti-conformity models}

To further understand deviations from the random-copying model, we next conducted an extension of the Slatkin exact test in which conformity or anti-conformity is directly incorporated in the null model (Methods). That is, instead of treating the probability that a non-mutant name is chosen as linearly proportional to its current frequency, we treat this probability as proportional to its frequency raised to a power $\beta$, with $\beta > 1$ representing conformity (preference for common variants) and $\beta < 1$ representing anti-conformity (preference for rare variants), and we evaluate the probability $P_E$ under a null model with parameter $\beta$.

In \cref{fig:biased test}A, we observe that for $\beta=0.5, 0.6, 0.7, 0.8$, and $0.9$, representing anti-conformity, $P_E$ always lies above $0.975$, indicating that the anti-conformity model is always rejected: the name sequence is more conformist than is specified by these parameter values. For $\beta=1.43$, 1.67, and 2, $0.025 < P_E < 0.975$ in the early years of the sequence, but $P_E < 0.025$ in the later years. That is, these values of $\beta$ provide a reasonable description of the conformity in the early years but are too conformist for recent years. In contrast, for $\beta=1.11$ and 1.25, $P_E > 0.975$ in the early years but $0.025 < P_E < 0.975$ in the later years, indicating that these values of $\beta$ are insufficiently conformist for the early years but are appropriate more recently.

Examining 200-year windows with conformist and anti-conformist null models (\cref{fig:biased test}B), we find that in most windows, anti-conformist models are rejected as insufficiently conformist ($P_E > 0.975$) and conformist models are not rejected ($0.025 < P_E < 0.975$). However, in the same periods in which the random-copying model produces dips in $P_E$ (\cref{fig:exact test}A), conformist but not anti-conformist models are rejected.

\section{Discussion}\label{sec:discussion}

\begin{figure}
    \centering
    \includegraphics[width=0.8\linewidth]{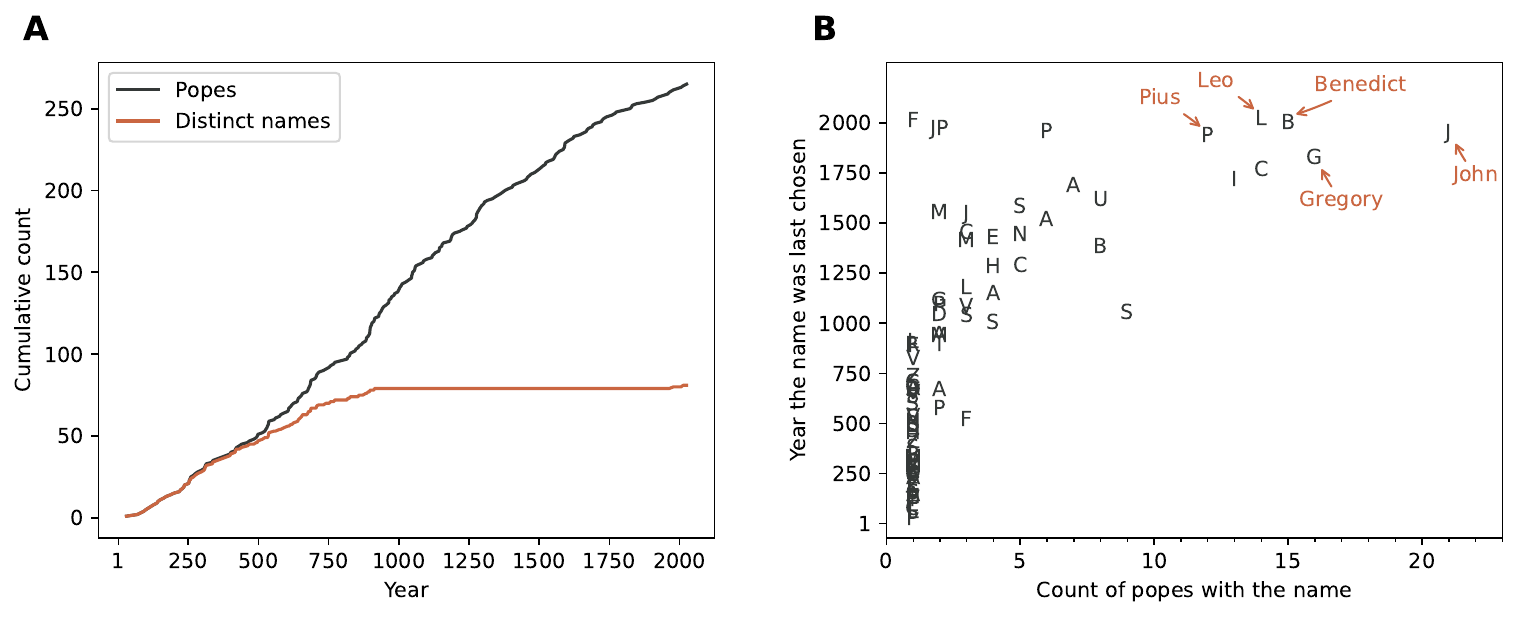}
    \caption{Papal names. (A) Cumulative counts of the number of popes and the number of distinct papal names. (B) Relationship between the year in which a name was last chosen and the number of occurrences of the name. Letters indicate the first letter of a papal name according to Anglophone spelling (``JP'' for the only two-word name, John Paul).}
    \label{fig:data} 
\end{figure}

We have examined the sequence of name choices made by Catholic popes (\cref{fig:data}) under a random-copying model for cultural evolution first developed in population genetics. Using the Slatkin exact test on the frequency vector of papal names, we found that the names often fit the random-copying model (\cref{fig:exact test}A); more frequently, the random-copying model is rejected with a $P$-value that corresponds to a conformist choice of names. In other words, although papal name choices are carefully considered, viewed from most points in time, the collection of sequential papal name choices amounts to randomly copying a name from among all previous popes, allowing for introduction of novel names, or to copying names with a supralinear conformity. The papal naming system provides an example of a cultural setting in which a large number of individual decisions, each likely deliberate and dependent on the context of the time, together generate an emergent pattern that conforms well to a simple mathematical model. 


For cultural phenomena that unfold over time, random-copying represents a natural simple model for a mechanism for choosing cultural variants. Random-copying has been examined as a model for a variety of cultural systems, including stylistic variation in ceramics \citep{neiman_stylistic_1995, shennan_ceramic_2001}, citations of patent applications \citep{bentley_random_2004}, and word frequencies in linguistic corpora \citep{reali_words_2010,bentley_population-level_2011}. Our study indicates that even across the long duration of one of the longest documented cultural processes, random-copying provides a suitable null model. 

Although we used random copying as a null model, the actual human process of name selection clearly has not always followed random-copying. Early popes served under their birth names, and the tradition of adopting a different papal name developed over time~\citep{vinne_papal_2006}. The first pope known to have not used his personal name was John II in 533, who was named Mercurius at birth~\citep{mcbrien_pocket_2006}, and some subsequent popes continued using their personal names after the tradition of adopting a papal name developed (after year 1000, only Adrian VI in 1522 and Marcellus II in 1555~\citep{mcbrien_pocket_2006}). Nevertheless, examining the entire span of 265 popes, the random-copying model often fits the name sequence.

\begin{figure}
    \centering
    \includegraphics[width=0.9\linewidth]{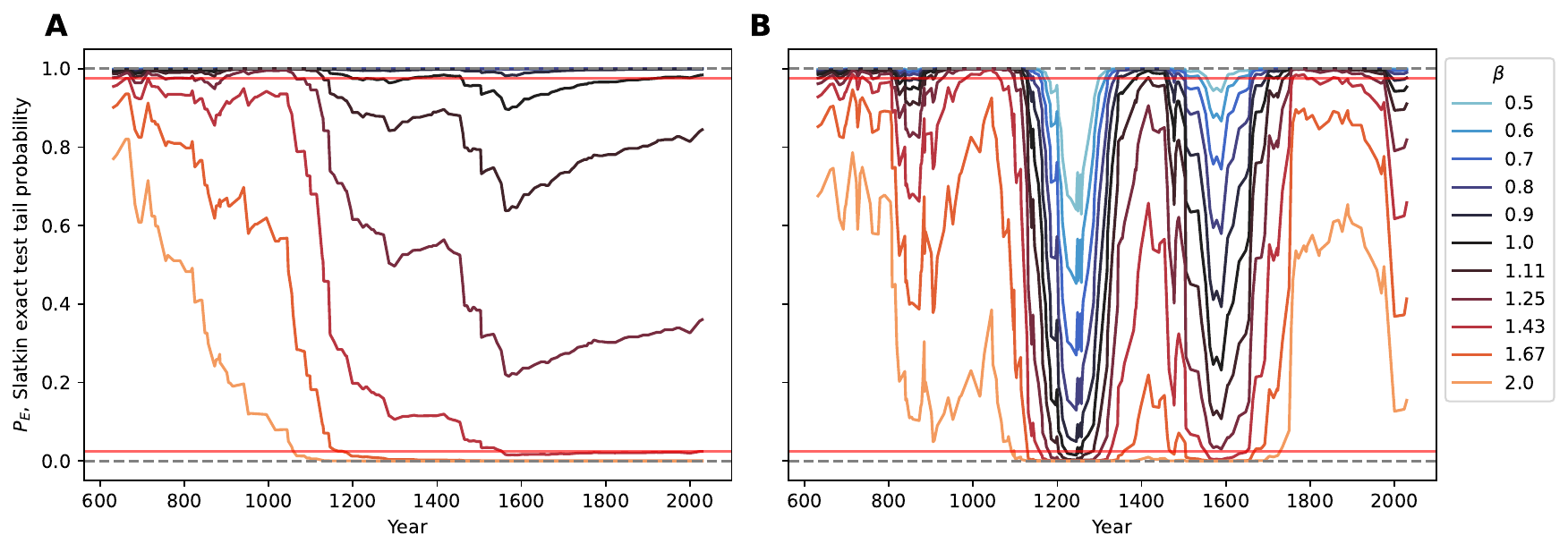}
    \caption{Slatkin exact test of null models with conformity or anti-conformity bias. The null model assumed is a model with a parameter $\beta$ representing conformity ($\beta > 1$) or anti-conformity ($\beta < 1$). (A) Slatkin exact test tail probability $P_E$. (B) Conditional Slatkin exact test tail probability $P_E$ for windows of width 200 years, indicated by the last year of the window. Red horizontal lines are placed at $P=0.025$ and $P=0.975$, representing the 5\% significance level of a two-tailed test.}
    \label{fig:biased test}
\end{figure}

While random-copying fits many temporal windows, conformity is preferred in most of the remaining windows. In the early years after the beginning of the adoption of papal names differing from birth names, relatively high values for the parameter $\beta$, representing conformity, are not rejected (\cref{fig:biased test}A). In the most recent hundreds of years, high values of conformity are rejected, but a more modest conformity fits the data.

We did identify periods in which neither random copying nor conformity fit closely (\cref{fig:window}A). Considering the diversity of papal names within windows (\cref{fig:window}C), during the 11th-13th centuries, rare but previously used names were chosen disproportionately often, at a \emph{higher} probability than expected under random-copying. Low values of $\beta$ are not rejected in windows centered on this period (\cref{fig:biased test}B). Such anti-conformity, or negative frequency-dependent bias \citep{boyd_culture_1988}, can be found in other quantitative cultural phenomena, such as baby name choices \citep{newberry_measuring_2022,acerbi_biases_2014} and the choice of chess moves at particular board positions \citep{lappo_cultural_2023}.

Writing in 1917 about papal onomastics, historian R.~L.~Poole~\cite{poole_names_1917} observed that from 1046 to 1144, 13 of 17 popes were the second of their name. The pattern continued: from 1145 to 1227, 8 of 12 popes were third of their name, and 1241 marks the beginning of a streak of five popes, each fourth of their name. The second-of-the-name streak came at a time in which popes might have sought to disassociate from a different period, during which the church was led mostly by Johns and Benedicts, and to instead reference older names associated with much earlier times~\citep{vinne_papal_2006}. 
 
To help understand how a long sequence of individual human decisions has produced a match to random copying or modest conformity, consider three alternatives. First, we can imagine a highly conformist scenario that would not fit random copying, for example, if at some stage, subsequent popes reused only a small number of names, repeatedly referencing the same few names and never referencing others. Indeed, the process might have been moving in this direction in the 11th century had the shift toward names of frequency 1, 2, and 3 not occurred. Second, we can also imagine an anti-conformist scenario that would not fit random copying, if the apparent preference for rare names evident in the 11th-13th centuries had continued. However, after incipient strong conformity was disrupted by anti-conformity, the anti-conformity dissipated and a closer match to random copying was restored. 

A third alternative is that the Slatkin exact test is underpowered to reject random copying with the number of popes to date. By adapting the random-copying model to incorporate a parameter $\beta$ that represents conformity or anti-conformity, we conducted a power analysis of the Slatkin exact test. This analysis suggests that enough popes are available to reject the stronger levels of conformity and anti-conformity that we have considered---and that neither strong conformity nor strong anti-conformity explains the data (\cref{sec:supp:power}).

Notably, we are treating the list of popes as definitive, using a list from the Catholic church itself~\citep{annuario}. At various times throughout church history, due to conflicts about who precisely was the true pope at particular moments, the list of previous popes would not have been as unambiguous as we treat it in our analysis---including to new popes choosing their papal names. Nevertheless, the name selections appear to reflect relatively simple, though changing, functions of the previous list of popes, even if some of the prior individuals now recognized as popes were not necessarily known to popes at the time of their name choices or were known but not recognized as popes. 

The random-copying model possesses a single parameter $\theta$ that in our context is related to the probability that a new, previously unseen name is chosen by a newly elected pope. We observed that maximum-likelihood estimates of $\theta$ decreased over time, both when inference was performed on the whole dataset (\cref{fig:exact test}B) and when it was performed in temporal windows (\cref{fig:window}B). This decrease in the ``mutation'' probability for papal names accords with the fact that for more than a millennium, from 914 to 1978, no papal election resulted in a previously unseen name. Even the 1978 mutation ``John Paul'' was not an entirely distinct name, as John Paul I chose the name to reference previous popes John XXIII and Paul VI~\citep{mcbrien_pocket_2006}.

The Slatkin exact test has been previously applied to cultural and archaeological data in attempts to detect possible deviations from random-copying \citep{steele_ceramic_2010, premo_cultural_2014, kandler_analysing_2019}. In the setting of testing the random-copying model, a novel aspect of our analysis is its application to an explicitly temporal process, providing a method for conducting the Slatkin exact test at a moment in time conditionally on the sequence of cultural choices in previous periods. A second novel aspect is that we used a biased Chinese restaurant process to enable conformity and anti-conformity models to be used as the null model, potentially enabling future tests of conformity and anti-conformity models in comparable discrete data sets.

The Ewens sampling formula and Chinese restaurant process have had broad applicability across probability and statistics and many data types, not only for population-genetic data and cultural data, but also for other types of data such as species abundances and word frequencies~\cite{crane_ubiquitous_2016, tavare_magical_2021}. Our conditional exact test and biased exact test (and biased conditional exact test) have potential to be adapted for application not only to other cultural data sets in which temporal ordering, conformity, and anti-conformity are of interest, but also to additional data sets in these other contexts.

\section{Methods}\label{sec:methods}

\subsection{Slatkin exact test}
\label{sec:methods:exact test}

Under the Ewens distribution describing a sample of alleles from a finite population \citep{ewens_sampling_1972}, the probability of a particular frequency vector of allelic types $\mathbf{c}=(c_1, c_2, \ldots, c_n)$ in a sample of size $n$ is given by 
\begin{equation}\label{eq:sampling formula}
    \probConfig{C_1=c_1, C_2=c_2,\ldots, C_n = c_n \mid n, \theta} = \frac{n!}{\theta_{(n)}} \prod_{j=1}^n \left(\frac \theta j\right)^{c_j} \frac{1}{c_j!},
\end{equation}
where $C_j$ is the random count of allelic types present in $j$ copies, $\theta$ is a mutation rate parameter, and $\theta_{(n)} = \theta (\theta + 1) \cdots (\theta + n - 1)$ is the rising factorial. Note that we require that $\sum_{j=1}^n j c_j = n$ in \cref{eq:sampling formula}. We include the subscript ``config'' to indicate that the argument $\mathbf{c}$ is a configuration that tabulates allele frequencies and does not contain information about the sequence in which alleles were sampled.

The random number $K$ of allelic types observed in the sample can be computed from the frequency vector as $K=\sum_{j=1}^n C_j$. For $1 \leq k \leq K$, the probability distribution of the number of allelic types is 
\begin{equation}\label{eq:k distribution} 
     \prob{K=k \mid n, \theta} = \stirling{n}{k}\frac{\theta^k}{\theta_{(n)}}, 
\end{equation}
where $\stirling{n}{k}$ is the unsigned Stirling number of the first kind \citep{ewens_sampling_1972,tavare_magical_2021}.

Conditioning the distribution in \cref{eq:sampling formula} on $K=k$, again requiring $\sum_{j=1}^n jc_j = n$, as well as $\sum_{j=1}^n c_j = k$,
\begin{equation}\label{eq:sampling fixed k}
 \probConfig{C_1=c_1, C_2=c_2, \ldots, C_n = c_n \mid n, \theta, k} = \frac{n!}{\stirling{n}{k}} \prod_{j=1}^n \left(\frac 1 j\right)^{c_j} \frac{1}{c_j!}.   
\end{equation}
\Cref{eq:sampling fixed k} is independent of $\theta$, so that $k$ is a sufficient statistic for $\theta$.

Slatkin's exact test uses the Ewens distribution to test the hypothesis that allele frequencies follow the infinitely-many-alleles model. More precisely, its null hypothesis is the statement ``the observed allele frequencies represent a realization of the infinitely-many-alleles model.'' Here, for fixed $k$ and $n$, each of the distinct realizations of the infinitely-many-alleles model is an ordered list $\mm =\left[a_1, a_2, \ldots, a_n\right]$ of $n$ sampled alleles, where each sampled allele has one of the $k$ allelic types, or $a_i \in \{1, 2, \ldots, k\}$. The allelic types are labeled in sequential order of their first appearance in the sequential list of draws.

In the context of papal names, $\mm$ corresponds to a particular temporal sequence of 265 named popes, distributed across 81 distinct names (Supplementary Table): ``Peter, Linus, Cletus, \ldots, Benedict, Francis, Leo''. This realization produces the frequency vector $\cc_0 = \left(44, 10, 0, 7, 4, 3, 2, 1, 2, 1, 0, 0, 1, 1, 2, 1, 1, 0, 0, 0, 0, 1\right)$.

The test statistic for Slatkin's exact test is the tail probability of the observed realization in the distribution of all possible realizations. In the Ewens sampling theory, the probability of a realization depends only on its frequency vector $\cc_0$ (in addition to fixed quantities $n$ and $k$). That is, \cref{eq:sampling fixed k} gives the sum of the probabilities of all realizations that share a frequency vector $\cc_0$. The number of such realizations is the number of ways that a sequence containing $c_1$ elements with frequency 1, $c_2$ elements with frequency 2, and so on, up to $c_n$ elements with frequency $n$ can be permuted, or $n! / (c_1! \, c_2! \cdots c_n!)$.

Therefore, via \cref{eq:sampling fixed k}, the probability of a particular realization $\mm$ of the infinitely-many-alleles model is given by \citep{ewens_sampling_1972}:
\begin{equation}
    \label{eq:test configuration prob}
    \probRealiz{C = \cc \mid n, \theta, k} = 
    \left. \bigg[ \frac{n!}{\stirling{n}{k}} \prod_{j=1}^n \left(\frac 1 j\right)^{c_j} \frac{1}{c_j!} \bigg] \middle / \frac{n!}{c_1! \, c_2! \cdots c_n!} \right.
    = \frac{1}{\stirling{n}{k}} \prod_{j=1}^{n} \bigg( \frac{1}{j} \bigg)^{c_j}.
\end{equation}
Subscript ``realiz'' indicates a computation of the probability of a particular realization that gives rise to configuration $\mathbf{c}$; all realizations that produce $\mathbf{c}$ are equiprobable. Note that \cite{slatkin_correction_1996} uses terms ``ordered configuration'' and ``unordered configuration'' for our concepts of ``configuration'' and ``realization,'' respectively.

The Slatkin exact test rejects the null hypothesis if the observed realization is unlikely to be observed under the Ewens sampling model.
The tail probability of the Slatkin exact test is evaluated as
\begin{equation}\label{eq:ewens tail probability}
    P_E = \sum_{i=1}^{D(n,k)}  \probConfig{\cc_i\mid n, k} \, \bigg\llbracket \probRealiz{\cc_i\mid n, k} \leq \probRealiz{\cc_0 \mid n, k} \bigg\rrbracket,
\end{equation}
where $\llbracket \cdot \rrbracket$ is the Iverson bracket, equal to 1 if the statement inside the brackets holds, and 0 otherwise \citep{slatkin_correction_1996}. The limit of summation $D(n,k)$ is the number of frequency vectors that have $\sum_{j=1}^n j c_j = n$ and $\sum_{j=1}^n c_j = k$, the number of partitions of $n$ into $k$ positive parts (OEIS A008284~\cite{oeis}). The $\probConfig{\cc_i\mid k}$ term weights realizations by their probability under the Ewens distribution; the use of the Iverson bracket tests if a sampled realization has probability less than or equal to that of the observed realization. Following \cite{slatkin_correction_1996}, the test is said to reject the null hypothesis at significance level $\alpha$ if $P_E < \alpha/2$ or $P_E > 1 - \alpha/2$.

\subsection{Implementation of the Slatkin exact test}\label{sec:methods:exact test implementation}

For large values of $n$ and $k$, the number $D(n,k)$ of possible frequency vectors with sample size $n$ and $k$ allelic types is large, and exhaustive enumeration of frequency vectors is impractical. The Slatkin exact test is then performed by simulation. Instead of summing the probabilities of vectors $\cc_i$ as in \cref{eq:ewens tail probability}, we generate a sample of frequency vectors from the Ewens distribution conditional on $n$ and $k$ (configurations) and then \emph{count} the fraction of vectors in the sample with probability under the model (\cref{eq:test configuration prob}) smaller than or equal to that of the observed $\cc_0$ \citep{slatkin_correction_1996}. We used $N=10^7$ simulated frequency vectors for each instance of a computation of the Slatkin exact test $P$-value. To avoid obtaining $P$-values of zero, we report $(a+1)/(N+1)$ as the $P$-value, where $a$ is the number of simulated frequency vectors with probability less than or equal to $\cc_0$.

An efficient algorithm for sampling frequency vectors from the Ewens distribution (\cref{eq:sampling formula}) is the Feller coupling algorithm~\citep{arratia_logarithmic_2003, tavare_magical_2021}, which, for fixed values of $n$ and $\theta$, obtains a frequency vector from Bernoulli random variables. In particular, the algorithm simulates Bernoulli random variables $\zeta_1, \zeta_2, \ldots, \zeta_n$, defined as
\begin{equation}\label{eq:zeta}
    \mathbb P[\zeta_i = 1] = \frac{\theta}{\theta + i - 1}.
\end{equation}
Next, the distances between consecutive 1's are identified in the sequence $\zeta_1, \zeta_2, \zeta_3, \ldots, \zeta_n, 1$ \citep{arratia_poisson_1992, tavare_magical_2021}, noting that it always holds that  $\zeta_1=1$. For example, if $n=7$ and the Bernoulli sequence $1, \zeta_2, \zeta_3, \ldots, \zeta_7, 1$ is $1, 0, 0, 1, 0, 1, 1, 1$, then the 1's are in positions 1, 4, 6, 7, and 8, the distances between the consecutive 1's are 3, 2, 1, and 1, and the sampled configuration is $\cc = (2, 1, 1)$. 

The Slatkin exact test requires sampling from the Ewens distribution \emph{conditional on the number of observed allelic types $k$} (\cref{eq:sampling fixed k}). In sampling via Feller coupling, a new allelic type is produced each time $\zeta_i = 1$ is observed, and hence, $\sum_{i=1}^n \zeta_i = k$. Therefore, sampling conditional on $k$ can be accomplished by sampling the Bernoulli random variables $\zeta_1, \zeta_2, \ldots, \zeta_n$ conditional on their sum being equal to $k$. To implement the sampling of independent but non-identically distributed Bernoulli random variables conditional on their having a fixed sum, we applied the procedure of \cite{heng_simple_2020}, extracting the sampled configuration $\cc$ from the distances between consecutive 1's.

Note that although in \cref{eq:sampling fixed k}, the Ewens distribution conditional on $k$ is independent of $\theta$, the Feller coupling approach requires a choice of $\theta$ to evaluate Bernoulli probabilities in \cref{eq:zeta}. This choice for $\theta$ does not affect the sampling or the results of the exact test, and we have arbitrarily chosen to set $\theta$ equal to its maximum likelihood estimate conditional on $n$ and $k$ (see \cref{sec:methods:theta inference}).

\subsection{Conditional exact test}
\label{sec:methods:conditional test}

To assess if papal name choices restricted to particular time periods conform to the random-copying model, we introduce an adjustment to the Slatkin exact test. The names in a sequence are separated into two types: first, the ``focal'' popes whose properties are tested for agreement with random copying, and, second, the predecessor popes, who are treated as fixed. 

For the conditional exact test, given an observed frequency vector $\cc_0$ with $n$ popes and $k$ distinct names, we identify the vector  $\cc_{(p)}$ constructed from the first $n_{(p)} < n$ popes and $k_{(p)} \leq k$ distinct names, representing fixed predecessor popes. In particular, dividing \cref{eq:test configuration prob} evaluated for an arbitrary vector $\cc$ with $n$ popes and $k$ names by \cref{eq:test configuration prob} evaluated for $\cc_{(p)}$, 
\begin{equation}\label{eq:conditional test p}
 \probRealiz{\cc \mid n, k, \cc_{(p)}} = \frac{\probRealiz{\cc\mid n, k}}{\probRealiz{\cc_{(p)} \mid n_{(p)}, k_{(p)}}} =  \frac{\stirling{n_{(p)}}{k_{(p)}}}{\stirling{n}{k}} \prod_{j=1}^{n} \bigg( \frac{1}{j} \bigg)^{c_j - c_{(p)j}}.
\end{equation}
For the $P$-value for the conditional exact test, we adjust \cref{eq:ewens tail probability} as follows:
\begin{equation}\label{eq:conditional tail probability}
    P_E = \sum_{i=1}^{D(n,k, \cc_{(p)})}  \probConfig{\cc_i\mid n, k, \cc_{(p)}} \, \bigg\llbracket \probRealiz{\cc_i\mid n, k, \cc_{(p)}} \leq \probRealiz{\cc_0 \mid n, k, \cc_{(p)}} \bigg\rrbracket.
\end{equation}
Here, $D(n,k, \cc_{(p)})$ is a certain subset of $D(n,k)$, with sample size $n$ and $k$ distinct names. In particular, $\cc_i$ is in $D(n,k, \cc_{(p)})$ if and only if each distinct name that is tabulated in the construction of the summary vector $\cc_i$ from a list of names with their multiplicities is observed at frequency greater than or equal its corresponding frequency in the construction of $\cc_{(p)}$.

As was true for the unconditional Slatkin exact test, for large $n$, the conditional exact test can be performed by stochastic sampling rather than enumeration of frequency vectors. Conditional sampling of frequency vectors can be implemented with the Chinese restaurant process \cite{tavare_magical_2021}. Given a realization of Bernoulli variables $\zeta_1, \zeta_2, \ldots, \zeta_n$ sampled as above (\cref{eq:zeta}), the Chinese restaurant process considers sampled alleles as ``customers'' at a restaurant, each assigned to a particular ``table,'' each of which corresponds to an allelic type. The Chinese restaurant process builds the frequency vector as follows: if $\zeta_i = 1$, then the $i$th customer is seated alone at a new table. If $\zeta_i = 0$, then the $i$th customer is seated at a table that is already populated, with the table selected in proportion to its number of people who have previously been seated.

To reflect conditioning on the vector $\cc_{(p)}$, the restaurant is treated as partially pre-filled according to $\cc_{(p)}$, with $n_{(p)}$ customers fixed at $k_{(p)}$ tables. Bernoulli random variables $\zeta_{n-n_{(p)}+1}, \zeta_{n-n_{(p)}+2}, \ldots, \zeta_{n}$ are sampled conditional on $\sum_{i=n_{(p)}+1}^n\zeta_{i} =  k - k_{(p)}$, and their values are used to seat the remaining $n-n_{(p)}$ customers. 

We used $10^7$ simulated frequency vectors for each computation of a conditional exact test. We implemented our algorithms for performing the unconditional and conditional exact tests a a command-line application written in the Rust programming language. The program is available at \href{https://github.com/EgorLappo/ewens-sampler}{\texttt{github.com/EgorLappo/ewens-sampler}}.

\subsection{Inference of \texorpdfstring{$\theta$}{θ}}\label{sec:methods:theta inference}

Given the number of popes $n$ and the number of distinct names $k$, we obtained the maximum likelihood estimate of $\theta$ from the likelihood function in \cref{eq:k distribution}. In particular, taking the logarithm of \cref{eq:k distribution}, we obtain
\begin{equation}
    f(\theta) = \log \prob{k\mid n, \theta} = \log \stirling{n}{k} + k \log \theta - \sum_{i=0}^{n-1} \log (\theta + i).
\end{equation}
We maximized the likelihood numerically.

Analogously to the extension of the Slatkin exact test to a conditional Slatkin exact test, we also computed conditional maximum likelihood estimates of $\theta$ given a fixed vector with sample size $n_{(p)} < n$ and $k_{(p)} \leq k$ types. The conditional likelihood function satisfies 
\begin{equation}
    \widetilde{f}(\theta) = \log \prob{k\mid n, n_{(p)}, k_{(p)}, \theta} = \log \left( \frac{\prob{k\mid n, \theta}}{\prob{k_{(p)}\mid n_{(p)}, \theta}} \right) \propto (k-k_{(p)}) \log \theta - \sum_{i=n_{(p)}}^{n-1} \log (\theta + i).
\end{equation}
We conducted the numerical maximization as above, specifying that in the special case of $k_{(p)}=k$, the maximum likelihood estimate is $\hat{\theta} = 0$.

\subsection{Diversity of names in a temporal window}\label{sec:methods:diversity}

To obtain the diversity of papal names within a temporal window, we evaluated an entropy statistic. Suppose that in a given 200-year window, a total of $n_w$ popes began their term, representing $k_w$ distinct names. Denoting the count of popes in the window with name $i$, $i=1,2,\ldots,k_w$, by $r_i$, entropy is computed as
\begin{equation}
    H = - \sum_{i=1}^{k_w} \frac{r_i}{n_w} \log\left(\frac{r_i}{n_w}\right).
\end{equation}

\subsection{Modeling conformity and anti-conformity}\label{sec:methods:biased}

Recall from \cref{sec:methods:conditional test} that if a customer in the Chinese restaurant process is placed at an already occupied table, the customer is seated at a table in proportion to its number of occupants. Mathematically, the $i$th customer chooses a pre-occupied table randomly with weights $p_1, p_2, \ldots, p_{k'}$ where $k' \leq k$ is the number of occupied tables, and $p_j$ is the fraction of the first $i-1$ customers sitting at the $j$th table.
However, one can imagine a \emph{biased} process in which customers prefer tables with relatively higher (conformity) or lower occupancy (anti-conformity). We formalize the biased process by introducing an additional parameter $\beta > 0$.

To simulate conformity or anti-conformity with parameter $\beta$, we proceed as follows. First, given values of $n$ and $k$ and possibly an initial configuration $\cc_0$, the variables $\zeta_i$ determining the sequence in which new tables are occupied are generated in the same manner as for the unbiased Chinese restaurant process. The $i$th customer then chooses an occupied table randomly with weights $p_1^\beta, p_2^\beta, \ldots, p_{k'}^\beta$. If $\beta = 1$, then we recover the usual unbiased process. If $\beta > 1$, because the function $x\mapsto x^\beta$ is convex, high-occupancy tables are chosen preferentially. Conversely, if $\beta < 1$, the function $x\mapsto x^\beta$ is concave, and low-occupancy tables are preferred.

In the context of papal name choices, $\beta > 1$ corresponds to popes preferring high-frequency names (conformity) and $\beta < 1$ corresponds to popes preferentially choosing rarely used names (anti-conformity). Note that the biased process does not affect the probability of innovation of novel names, only the probabilities of the various previously chosen names conditional on the choice to use a previously used name.

The simulation-based Slatkin exact test can be straightforwardly adapted to test if data accord with the biased model. Given the observed frequency vector $\cc_0$, a sample of frequency vectors is generated from the biased process with a specified value of $\beta$, and the tail probability is obtained as the fraction of sampled vectors with probability (\cref{eq:test configuration prob}) lower than or equal to that of $\cc_0$. We performed tests based on the biased Chinese restaurant process using the same approach as in \cref{sec:methods:exact test implementation} above, simulating $N=10^5$ configurations.

\vskip .3cm
{\small
\noindent \textbf{Data and code availability.} Code to perform the analyses in this paper and to replicate the presented figures is available online at \href{https://github.com/EgorLappo/papal-names}{\texttt{github.com/EgorLappo/papal-names}}. Our implementation of the Slatkin exact test is available as a command-line program at \href{https://github.com/EgorLappo/ewens-sampler}{\texttt{github.com/EgorLappo/ewens-sampler}}.

\vskip .3cm
\noindent \textbf{Acknowledgments.} We thank N.~Egan for helpful conversations and two reviewers for comments on the manuscript.

\vskip .3cm
\noindent \textbf{Declaration of interests.} The authors declare no competing interests.
}
\printbibliography
\clearpage

\clearpage

\clearpage
\begin{figure}
    \centering
    \includegraphics[width=0.8\linewidth]{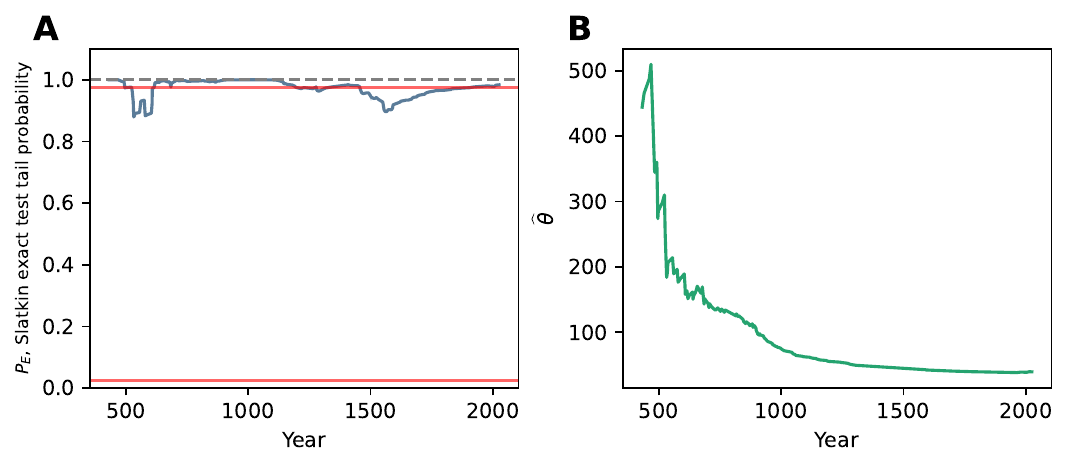}
    \caption{Slatkin exact test $P$-value and estimated parameter value $\hat\theta$ for the fit of papal name choices to a random-copying model. For each pope starting with Sixtus III in year 432---the first pope for which the name frequency distribution conditional on the number of names is nontrivial---the list of popes is truncated, the $P$-value is computed, and the parameter $\hat\theta$ is estimated. The $x$-axis values correspond to the first year of the term for the last pope included in the truncated dataset. (A) Slatkin exact test tail probability $P_E$. Red horizontal lines are placed at $P=0.025$ and $P=0.975$, representing the 5\% significance level of a two-tailed test. (B) Maximum likelihood estimate of the scaled population mutation rate $\theta$. }
    \label{fig:exact test}
\end{figure}

\clearpage
\begin{figure}
    \centering
    \includegraphics[width=0.9\linewidth]{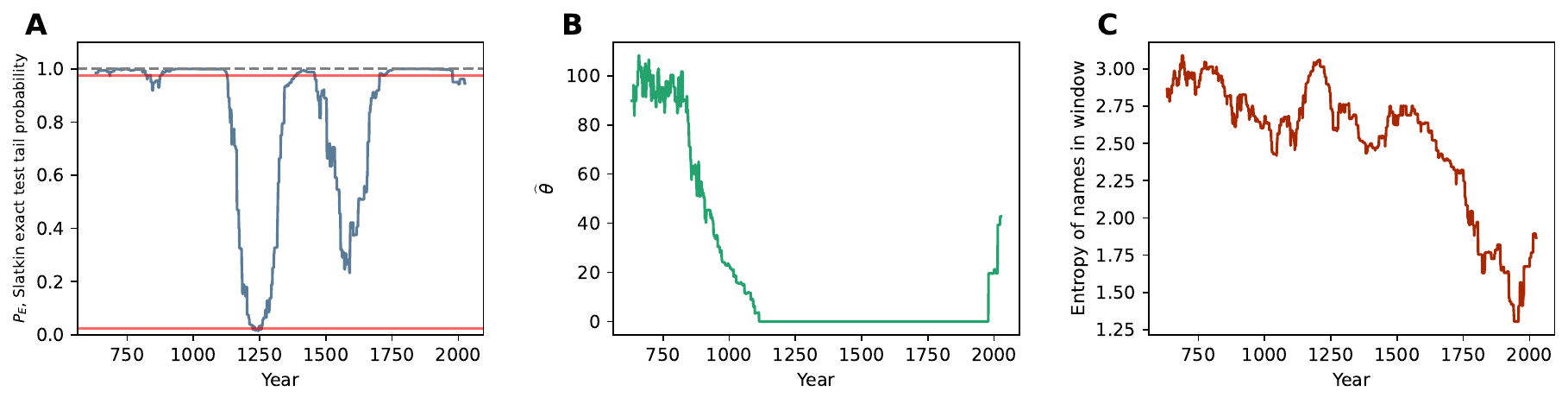}
    \caption{Slatkin exact test $P$-value, estimated parameter value $\hat\theta$, and entropy computed for sliding windows. For windows of width 200 years, indicated by the last year of the window, only popes who began their term during the window are considered. (A) Conditional Slatkin exact test tail probability $P_E$. Red horizontal lines are placed at $P=0.025$ and $P=0.975$, representing the 5\% significance level of a two-tailed test. (B) Maximum likelihood estimate of the scaled population mutation rate $\theta$ for the window. (C) Shannon entropy of the set of papal names within the window.}
    \label{fig:window}
\end{figure}

\clearpage

\clearpage

\noindent\textbf{\LARGE Supplementary information}

\appendix

\numberwithin{equation}{section}
\setcounter{figure}{0}
\renewcommand\thefigure{S\arabic{figure}}
\renewcommand\thesection{S\arabic{section}}
\renewcommand\thetable{S\arabic{table}}

\section{Power analysis}\label{sec:supp:power}

We analyzed the power of the Slatkin exact test to reject a null hypothesis of neutrality. For this power analysis, we simulated alternative hypotheses with biased configurations, using the model of \cref{sec:methods:biased} and values of the bias parameter $\beta$ ranging from $0.5$ (strong anti-conformity) to 2 (strong conformity).

\begin{figure}[bh!]
    \centering
    \includegraphics[width=0.9\linewidth]{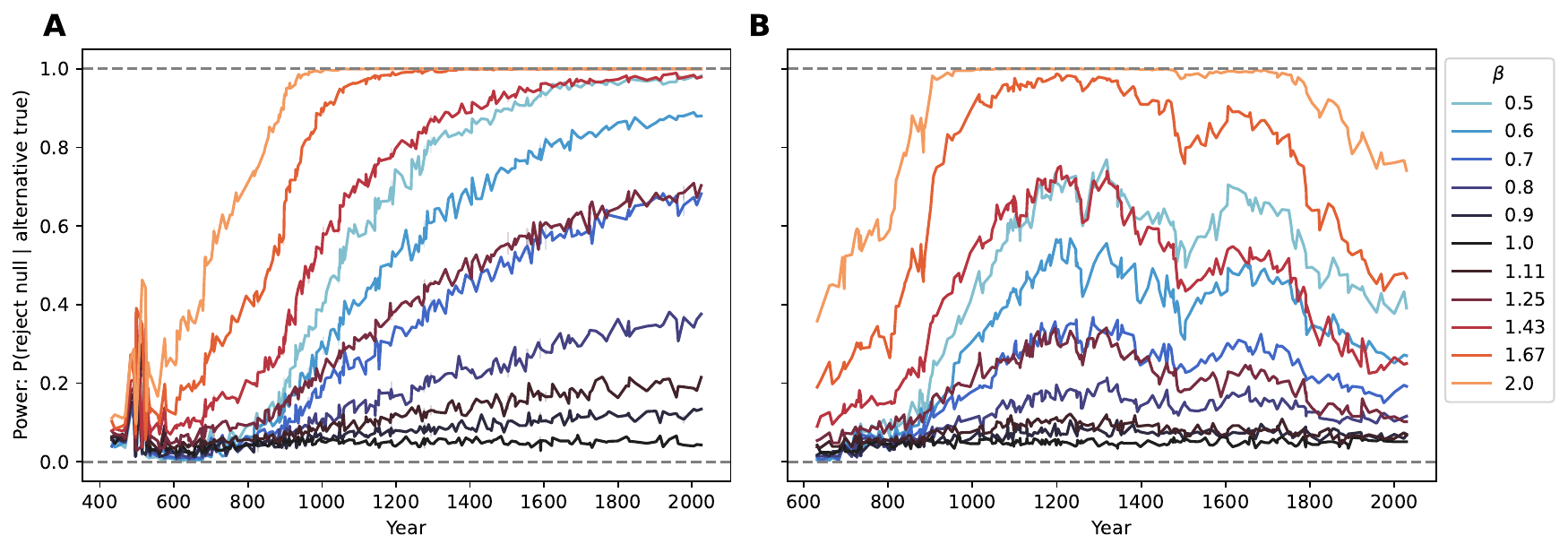}
    \caption{Statistical power of the Slatkin exact test to reject the null model of random copying ($\beta=1$). (A) Statistical power of the Slatkin exact test. (B) Statistical power of the conditional Slatkin exact test for windows of width 200 years, indicated by the last year of the window. Power to reject the null hypothesis $\beta=1$ is evaluated using simulations with values $\beta \neq 1$ that represent alternative hypotheses.}
    \label{fig:power}
\end{figure}

To be precise, recalling that the statistical power of a test is defined as the probability of rejecting the null hypothesis given that the alternative hypothesis is true, in our case, the null hypothesis is neutrality, $\beta = 1$, and the alternative hypothesis is $\beta$ equal to some other value. For a power analysis over the entire duration of the data set up through a specific pope $i$ (as in \cref{fig:exact test}A), we recorded the number of popes $n_i$ and number of unique names $k_i$ in configuration $\cc_i$ at each pope $i$. For each $(n_i,k_i)$, we then simulated 1000 frequency vectors for each value of $\beta$ in $\{0.5, 0.6, 0.7, 0.8, 0.9, 1.0, 1.11, 1.25, 1.43, 1.67, 2.0\}$. For each of tbe 1000 vectors at a specified $\beta$, for each $i$, we conducted the Slatkin exact test (simulating $10^5$ configurations for conducting the test). In other words, we computed the $P$-value for each of the 1000 vectors, tabulating the fraction of $P$-values that resulted in rejection of the null hypothesis at the 0.05 significance level  (\cref{fig:power}A).

For a power analysis of the conditional exact test, we proceeded similarly, mimicking \cref{fig:window}A. For each 200-year moving window $i$, we recorded the frequency vector of papal names at the beginning of the window $\cc_{(p), i}$, the number of popes at the end of the window $n_i$, and the number of unique names observed at the end of the window $k_i$. For each $\beta$, we then simulated 1000 frequency vectors with $n_i$ popes and $k_i$ names, all starting with the same fixed configuration $\cc_{(p),i}$. We tested each simulated configuration for accordance with the neutral model using a conditional exact test with initial vector $\cc_{(p),i}$ and $10^5$ simulated configurations, then recorded the fraction of tests that led to rejection at the 0.05 significance level (\cref{fig:power}B).

The results of our power computations demonstrate that by the end of the 2000-year long process, the Slatkin exact test has relatively high power to reject strong conformity ($\beta=1.25, 1.43, 1.67, 2.0$) and strong anti-conformity ($\beta=0.5, 0.6, 0.7$) in papal name choices (\cref{fig:power}A); and power was low only for the weakest levels of conformity ($\beta=1.11$) and anti-conformity ($\beta=0.8, 0.9$). Note that the power to reject the null at $\beta=1$ is near 0.05, as expected with use of a 0.05 significance threshold.

For the conditional test in windows, because windows have fewer popes than the full 2000-year period, power is generally not as high as for the full process. Nevertheless, the test generally has enough power to reject the strongest values of conformity and anti-conformity (\cref{fig:power}B).

\clearpage 

\section{List of popes}\label{sec:supp:list}

\begin{multicols*}{3}
\TrickSupertabularIntoMulticols
\tablehead{\toprule & Pope & Year \\ \midrule}
\tabletail{\bottomrule}
\bottomcaption{List of popes. In the table, year denotes the year of election, ``*'' denotes popes who popes who were the first to choose their name, and ``\textdagger'' denotes popes who are (currently) the last to choose their name. Wherever the Annuario Pontificio \citep{annuario} gives the election years as uncertain, we choose the earliest year.}\label{table:popes}
\begin{supertabular}{llr}
1 & Peter\textsuperscript{*}\textsuperscript{\textdagger} & 30 \\
2 & Linus\textsuperscript{*}\textsuperscript{\textdagger} & 68 \\
3 & Cletus\textsuperscript{*}\textsuperscript{\textdagger} & 80\\
4 & Clement I\textsuperscript{*} & 92 \\
5 & Evaristus\textsuperscript{*}\textsuperscript{\textdagger} & 99 \\
6 & Alexander I\textsuperscript{*} & 108 \\
7 & Sixtus I\textsuperscript{*} & 117 \\
8 & Telesphorus\textsuperscript{*}\textsuperscript{\textdagger} & 127 \\
9 & Hyginus\textsuperscript{*}\textsuperscript{\textdagger} & 138 \\
10 & Pius I\textsuperscript{*} & 142 \\
11 & Anicetus\textsuperscript{*}\textsuperscript{\textdagger} & 150 \\
12 & Soter\textsuperscript{*}\textsuperscript{\textdagger} & 162 \\
13 & Eleutherius\textsuperscript{*}\textsuperscript{\textdagger} & 171 \\
14 & Victor I\textsuperscript{*} & 186 \\
15 & Zephyrinus\textsuperscript{*}\textsuperscript{\textdagger} & 198 \\
16 & Callixtus I\textsuperscript{*} & 218 \\
17 & Urban I\textsuperscript{*} & 222 \\
18 & Pontian\textsuperscript{*}\textsuperscript{\textdagger} & 230 \\
19 & Anterus\textsuperscript{*}\textsuperscript{\textdagger} & 235 \\
20 & Fabian\textsuperscript{*}\textsuperscript{\textdagger} & 236 \\
21 & Cornelius\textsuperscript{*}\textsuperscript{\textdagger} & 251 \\
22 & Lucius I\textsuperscript{*} & 253 \\
23 & Stephen I\textsuperscript{*} & 254 \\
24 & Sixtus II & 257 \\
25 & Dionysius\textsuperscript{*}\textsuperscript{\textdagger} & 259 \\
26 & Felix I\textsuperscript{*} & 269 \\
27 & Eutychian\textsuperscript{*}\textsuperscript{\textdagger} & 275 \\
28 & Caius\textsuperscript{*}\textsuperscript{\textdagger} & 283 \\
29 & Marcellinus\textsuperscript{*}\textsuperscript{\textdagger} & 296 \\
30 & Marcellus I\textsuperscript{*} & 306 \\
31 & Eusebius\textsuperscript{*}\textsuperscript{\textdagger} & 309 \\
32 & Miltiades\textsuperscript{*}\textsuperscript{\textdagger} & 311 \\
33 & Sylvester I\textsuperscript{*} & 314 \\
34 & Mark\textsuperscript{*}\textsuperscript{\textdagger} & 336 \\
35 & Julius I\textsuperscript{*} & 337 \\
36 & Liberius\textsuperscript{*}\textsuperscript{\textdagger} & 352 \\
37 & Damasus I\textsuperscript{*} & 366 \\
38 & Siricius\textsuperscript{*}\textsuperscript{\textdagger} & 384 \\
39 & Anastasius I\textsuperscript{*} & 399 \\
40 & Innocent I\textsuperscript{*} & 401 \\
41 & Zosiumus\textsuperscript{*}\textsuperscript{\textdagger} & 417 \\
42 & Boniface I\textsuperscript{*} & 418 \\
43 & Celestine I\textsuperscript{*} & 422 \\
44 & Sixtus III & 432 \\
45 & Leo I\textsuperscript{*} & 440 \\
46 & Hilarius\textsuperscript{*}\textsuperscript{\textdagger} & 461 \\
47 & Simplicius\textsuperscript{*}\textsuperscript{\textdagger} & 468 \\
48 & Felix III & 483 \\
49 & Gelasius I\textsuperscript{*} & 492 \\
50 & Anastasius II & 496 \\
51 & Symmachus\textsuperscript{*}\textsuperscript{\textdagger} & 498 \\
52 & Hormisdas\textsuperscript{*}\textsuperscript{\textdagger} & 514 \\
53 & John I\textsuperscript{*} & 523 \\
54 & Felix IV\textsuperscript{\textdagger} & 526 \\
55 & Boniface II & 530 \\
56 & John II & 532 \\
57 & Agapetus I\textsuperscript{*} & 535 \\
58 & Silverius\textsuperscript{*}\textsuperscript{\textdagger} & 536 \\
59 & Vigilius\textsuperscript{*}\textsuperscript{\textdagger} & 537 \\
60 & Pelagius I\textsuperscript{*} & 556 \\
61 & John III & 561 \\
62 & Benedict I\textsuperscript{*} & 575 \\
63 & Pelagius II\textsuperscript{\textdagger} & 579 \\
64 & Gregory I\textsuperscript{*} & 590 \\
65 & Sabinian\textsuperscript{*}\textsuperscript{\textdagger} & 604 \\
66 & Boniface III & 607 \\
67 & Boniface IV & 608 \\
68 & Adeodatus I\textsuperscript{*} & 615 \\
69 & Boniface V & 619 \\
70 & Honorius I\textsuperscript{*} & 625 \\
71 & Severinus\textsuperscript{*}\textsuperscript{\textdagger} & 638 \\
72 & John IV & 640 \\
73 & Theodore I\textsuperscript{*} & 642 \\
74 & Martin I\textsuperscript{*} & 649 \\
75 & Eugene I\textsuperscript{*} & 654 \\
76 & Vitalian\textsuperscript{*}\textsuperscript{\textdagger} & 657 \\
77 & Adeodatus II\textsuperscript{\textdagger} & 672 \\
78 & Donus\textsuperscript{*}\textsuperscript{\textdagger} & 676 \\
79 & Agatho\textsuperscript{*}\textsuperscript{\textdagger} & 678 \\
80 & Leo II & 681 \\
81 & Benedict II & 684 \\
82 & John V & 685 \\
83 & Conon\textsuperscript{*}\textsuperscript{\textdagger} & 686 \\
84 & Sergius I\textsuperscript{*} & 687 \\
85 & John VI & 701 \\
86 & John VII & 705 \\
87 & Sisinnius\textsuperscript{*}\textsuperscript{\textdagger} & 708 \\
88 & Constantine\textsuperscript{*}\textsuperscript{\textdagger} & 708 \\
89 & Gregory II & 715 \\
90 & Gregory III & 731 \\
91 & Zachary\textsuperscript{*}\textsuperscript{\textdagger} & 741 \\
92 & Stephen II & 752 \\
93 & Paul I\textsuperscript{*} & 757 \\
94 & Stephen III & 768 \\
95 & Adrian\textsuperscript{*} & 772 \\
96 & Leo III & 795 \\
97 & Stephen IV & 816 \\
98 & Paschal I\textsuperscript{*} & 817 \\
99 & Eugene II & 824 \\
100 & Valentine\textsuperscript{*}\textsuperscript{\textdagger} & 827 \\
101 & Gregory IV & 827 \\
102 & Sergius II & 844 \\
103 & Leo IV & 847 \\
104 & Benedict III & 855 \\
105 & Nicholas I\textsuperscript{*} & 858 \\
106 & Adrian II & 867 \\
107 & John VIII & 872 \\
108 & Marinus I\textsuperscript{*} & 882 \\
109 & Adrian III & 884 \\
110 & Stephen V & 885 \\
111 & Formosus\textsuperscript{*}\textsuperscript{\textdagger} & 891 \\
112 & Boniface VI & 896 \\
113 & Stephen VI & 896 \\
114 & Romanus\textsuperscript{*}\textsuperscript{\textdagger} & 897 \\
115 & Theodore II\textsuperscript{\textdagger} & 897 \\
116 & John IX & 898 \\
117 & Benedict IV & 900 \\
118 & Leo V & 903 \\
119 & Sergius III & 904 \\
120 & Anastasius III & 911 \\
121 & Lando\textsuperscript{*}\textsuperscript{\textdagger} & 913 \\
122 & John X & 914 \\
123 & Leo VI & 928 \\
124 & Stephen VII & 929 \\
125 & John XI & 931 \\
126 & Leo VII & 936 \\
127 & Stephen VIII & 939 \\
128 & Marinus II\textsuperscript{\textdagger} & 942 \\
129 & Agapetus II\textsuperscript{\textdagger} & 946 \\
130 & John XII & 955 \\
131 & Benedict V & 964 \\
132 & Leo VIII & 964 \\
133 & John XIII & 965 \\
134 & Benedict VI & 972 \\
135 & Benedict VII & 974 \\
136 & John XIV & 983 \\
137 & John XV & 985 \\
138 & Gregory V & 996 \\
139 & Sylvester II & 999 \\
140 & John XVII & 1003 \\
141 & John XVIII & 1003 \\
142 & Sergius IV\textsuperscript{\textdagger} & 1009 \\
143 & Benedict VIII & 1012 \\
144 & John XIX & 1024 \\
145 & Benedict IX & 1032 \\
146 & Sylvester III\textsuperscript{\textdagger} & 1045 \\
147 & Gregory VI & 1045 \\
148 & Clement II & 1046 \\
149 & Damasus II\textsuperscript{\textdagger} & 1048 \\
150 & Leo IX & 1049 \\
151 & Victor II & 1055 \\
152 & Stephen IX\textsuperscript{\textdagger} & 1057 \\
153 & Nicholas II & 1058 \\
154 & Alexander II & 1061 \\
155 & Gregory VII & 1073 \\
156 & Victor III\textsuperscript{\textdagger} & 1086 \\
157 & Urban II & 1088 \\
158 & Paschal II\textsuperscript{\textdagger} & 1099 \\
159 & Gelasius II\textsuperscript{\textdagger} & 1118 \\
160 & Callixtus II & 1119 \\
161 & Honorius II & 1124 \\
162 & Innocent II & 1130 \\
163 & Celestine II & 1143 \\
164 & Lucius II & 1144 \\
165 & Eugene III & 1145 \\
166 & Anastasius IV\textsuperscript{\textdagger} & 1153 \\
167 & Adrian IV & 1154 \\
168 & Alexander III & 1159 \\
169 & Lucius III\textsuperscript{\textdagger} & 1181 \\
170 & Urban III & 1185 \\
171 & Gregory VIII & 1187 \\
172 & Clement III & 1187 \\
173 & Celestine III & 1191 \\
174 & Innocent III & 1198 \\
175 & Honorius III & 1216 \\
176 & Gregory IX & 1227 \\
177 & Celestine IV & 1241 \\
178 & Innocent IV & 1243 \\
179 & Alexander IV & 1254 \\
180 & Urban IV & 1261 \\
181 & Clement IV & 1265 \\
182 & Gregory X & 1271 \\
183 & Innocent V & 1276 \\
184 & Adrian V & 1276 \\
185 & John XXI & 1277 \\
186 & Nicholas III & 1277 \\
187 & Martin IV & 1281 \\
188 & Honorius IV\textsuperscript{\textdagger} & 1285 \\
189 & Nicholas IV & 1288 \\
190 & Celestine V\textsuperscript{\textdagger} & 1294 \\
191 & Boniface VIII & 1294 \\
192 & Benedict XI & 1303 \\
193 & Clement V & 1305 \\
194 & John XXII & 1316 \\
195 & Benedict XII & 1334 \\
196 & Clement VI & 1342 \\
197 & Innocent VI & 1352 \\
198 & Urban V & 1362 \\
199 & Gregory XI & 1370 \\
200 & Urban VI & 1378 \\
201 & Boniface IX\textsuperscript{\textdagger} & 1389 \\
202 & Innocent VII & 1404 \\
203 & Gregory XII & 1406 \\
204 & Martin V\textsuperscript{\textdagger} & 1417 \\
205 & Eugene IV\textsuperscript{\textdagger} & 1431 \\
206 & Nicholas V\textsuperscript{\textdagger} & 1447 \\
207 & Callixtus III\textsuperscript{\textdagger} & 1455 \\
208 & Pius II & 1458 \\
209 & Paul II & 1464 \\
210 & Sixtus IV & 1471 \\
211 & Innocent VIII & 1484 \\
212 & Alexander VI & 1492 \\
213 & Pius III & 1503 \\
214 & Julius II & 1503 \\
215 & Leo X & 1513 \\
216 & Adrian VI\textsuperscript{\textdagger} & 1522 \\
217 & Clement VII & 1523 \\
218 & Paul III & 1534 \\
219 & Julius III\textsuperscript{\textdagger} & 1550 \\
220 & Marcellus II\textsuperscript{\textdagger} & 1555 \\
221 & Paul IV & 1555 \\
222 & Pius IV & 1559 \\
223 & Pius V & 1566 \\
224 & Gregory XIII & 1572 \\
225 & Sixtus V\textsuperscript{\textdagger} & 1585 \\
226 & Urban VII & 1590 \\
227 & Gregory XIV & 1590 \\
228 & Innocent IX & 1591 \\
229 & Clement VIII & 1592 \\
230 & Leo XI & 1605 \\
231 & Paul V & 1605 \\
232 & Gregory XV & 1621 \\
233 & Urban VIII\textsuperscript{\textdagger} & 1623 \\
234 & Innocent X & 1644 \\
235 & Alexander VII & 1655 \\
236 & Clement IX & 1667 \\
237 & Clement X & 1670 \\
238 & Innocent XI & 1676 \\
239 & Alexander VIII\textsuperscript{\textdagger} & 1689 \\
240 & Innocent XII & 1691 \\
241 & Clement XI & 1700 \\
242 & Innocent XIII\textsuperscript{\textdagger} & 1721 \\
243 & Benedict XIII & 1724 \\
244 & Clement XII & 1730 \\
245 & Benedict XIV & 1740 \\
246 & Clement XIII & 1758 \\
247 & Clement XIV\textsuperscript{\textdagger} & 1769 \\
248 & Pius VI & 1775 \\
249 & Pius VII & 1800 \\
250 & Leo XII & 1823 \\
251 & Pius VIII & 1829 \\
252 & Gregory XVI\textsuperscript{\textdagger} & 1831 \\
253 & Pius IX & 1846 \\
254 & Leo XIII & 1878 \\
255 & Pius X & 1903 \\
256 & Benedict XV & 1914 \\
257 & Pius XI & 1922 \\
258 & Pius XII\textsuperscript{\textdagger} & 1939 \\
259 & John XXIII\textsuperscript{\textdagger} & 1958 \\
260 & Paul VI\textsuperscript{\textdagger} & 1963 \\
261 & John Paul I\textsuperscript{*} & 1978 \\
262 & John Paul II\textsuperscript{\textdagger} & 1978 \\
263 & Benedict XVI\textsuperscript{\textdagger} & 2005 \\
264 & Francis\textsuperscript{*}\textsuperscript{\textdagger} & 2013 \\
265 & Leo XIV\textsuperscript{\textdagger} & 2025 \\\end{supertabular}
\end{multicols*}

\end{document}